\documentstyle {article}
\font\script=cmmib10
\font\bbold=cmmib10

\newcommand{\SM}{\mbox{\script M}}
\newcommand{\BR}{\mbox{\bbold R}}
\begin{document}
\textwidth 18.0cm
\textheight 23.0cm
\topmargin -0.5in
\baselineskip 16pt
\parskip 18pt
\parindent 30pt
\title{ \large \bf A note on the averaged null energy condition
in quantum field theory}
\author{Ulvi Yurtsever \\
{}~~~~~~~~~~~~ \\
Jet Propulsion Laboratory 169-327\\
California Institute of Technology\\
4800 Oak Grove Drive\\
Pasadena, CA 91109\\
and\\
Theoretical Astrophysics 130-33\\
California Institute of Technology\\
Pasadena, CA 91125}
\date{January, 1995}
\pagestyle{empty}
\baselineskip 24pt
\maketitle
\thispagestyle{empty}
\vspace{.2in}
\baselineskip 12pt
\begin{abstract}
\noindent Locally, the stress-energy of quantized matter
can become arbitrarily
negative in a wide class of quantum states, thereby violating the
classical positive-energy conditions of General Relativity without bound.
Since without such constraints
the theory would have no predictive power,
uncovering what nonlocal constraints, if any, quantum field theory
imposes on the renormalized stress-energy tensor is of central
importance for semiclassical gravity. One such nonlocal constraint, the
averaged null energy condition (ANEC---the condition that the null-null
component of the stress-energy tensor integrated along a complete null
geodesic is nonnegative in every quantum state) has been recently
shown to hold for linear quantum fields in a large class of spacetimes.
Nevertheless, it is easy to show by using a simple
scaling argument that ANEC as stated cannot
hold generically in curved four-dimensional spacetime, and this
scaling argument has been widely
interpreted as a death-blow for averaged energy
conditions in quantum field theory. In this note I propose a simple
generalization of ANEC, in which the right-hand-side of the ANEC
inequality is replaced by a finite (but in general negative)
state-independent lower bound. As long as attention is focused
on asymptotically well-behaved spacetimes, this generalized version of
ANEC is safe from the threat of the scaling argument, and thus stands a
chance of being generally valid in four-dimensional curved spacetime.
I argue that when generalized ANEC holds, it has implications for the
non-negativity of total energy and for singularity theorems similar to
the implications of ANEC. In particular, I show that if generalized ANEC
is satisfied in static traversable wormhole spacetimes (which is likely
but remains to be shown), then macroscopic
wormholes (but not necessarily microscopic, Planck-size wormholes) are
ruled out by quantum field theory.
\vspace{0.5cm}

\end{abstract}
\newpage
\baselineskip 14pt
\parskip 10pt
\pagestyle{plain}
\pagenumbering{arabic}
{}~~~~~~

The most striking aspect of the violation of positive-energy conditions
by quantum stress-energy tensors is the unbounded extent of the
violation. For example, even for a Klein-Gordon
scalar field in flat, Minkowski space, the regularized
(normal-ordered)
expectation value $\langle \omega | \!\! : \!\! T_{00}(x) \!\! : \!\!
| \omega \rangle$ at any point $x$ is unbounded from below as a
functional of the quantum state $\omega$. Furthermore,
the volume integral of
$\langle \omega | \!\! : \!\! T_{00}(x) \!\! : \!\! | \omega \rangle$
over any fixed, spacelike 3-box of {\sl finite} size
is also unbounded from below
as a functional of $\omega$ (and a similar result holds for the
spacetime-volume integral over a compact 4-box; see
Sect.\,1 and Ref.\,[1] of [1] for more details). Given this tendency
of the regularized expectation value
$\langle \omega | T_{ab}(x) | \omega \rangle$
to become unboundedly negative, any condition that sets a lower bound
on nonlocal averages of
$\langle \omega | T_{ab}(x) | \omega \rangle$ would be a significant
constraint on the quantum stress-energy tensor.

In this note I propose the following
constraint as a generalization of the averaged null energy condition
(ANEC; see [2] and [1] for a discussion of ANEC and its brief
history): Let $\langle \omega | T_{ab} | \omega \rangle$ denote the
(renormalized) stress-energy tensor of a quantum field on a curved
spacetime $( \SM ,g)$, and let $\gamma \subset \SM$ be a complete null
geodesic. For $k^a$ a given (parallel-propagated) tangent
vector along $\gamma$, let me
introduce the following quantity $\beta (k)$:
\begin{equation}
\beta (k) \equiv \inf_{\omega} \int_{\gamma} \langle \omega
| T_{ab} | \omega \rangle \, k^a
k^b \, dv \; .
\end{equation}
I will say that $\langle T_{ab} \rangle$ satisfies generalized ANEC
along $\gamma$ if $\beta (k) > - \infty$. Here the infimum is taken
over all Hadamard states $\omega$ of the quantum
field, and the integral along
$\gamma$ is with respect to the affine
parameter $v$ which corresponds to the
tangent vector $k^a$ (i.e., $d{\gamma}^a /dv = k^a$). Properly interpreted,
the quantity $\beta$ is a
1-form, whose contraction with the tangent vector $k^a$ is given by the
right hand side of Eq.\,(1). More precisely, $\beta$ is an element of
the quotient space ${T_p}^{\ast}\SM / N_p$,
where $p$ is a point on $\gamma$,
$N_p \subset {T_p}^{\ast}\SM$ is the subspace of all 1-forms $\alpha
\in {T_p}^{\ast}\SM$ which annihilate the
tangent vector $\gamma_{\ast}$ [$\alpha
(\gamma_{\ast})=0$], and $\beta$ is parallel transported along $\gamma$ so
that it does not matter at which $p \in \gamma$ the quantity $\beta
(\gamma_{\ast})$ is evaluated.

The usual ANEC along $\gamma$ is recovered
by setting $\beta (\gamma_\ast ) \geq 0$.
If the integrand on the right hand side of Eq.\,(1) is non-integrable
for some Hadamard states $\omega$,
a more precise version of generalized ANEC needs to be employed just like
the more precise version of ANEC discussed in Sect.\,2 of [2]. Namely,
let $c(x)$ be a compact-supported real-valued
function on $\BR$ whose Fourier transform $\hat{c} (s)$ is such that
for some $\delta > 0$ the function
\mbox{$(1+s^2)^{1+ \delta} |\hat{c}(s)|$}
is bounded [which implies that $c(x)$ is $C^1$]. Generalized
ANEC holds along a complete null geodesic
$\gamma$ if for every such weighting
function $c(x)$ the 1-form
$\beta_c$ along $\gamma$ defined by
\begin{equation}
\beta_c (k ) \equiv \inf_{\omega} \; \liminf_{\lambda \rightarrow \infty}
\int_{\gamma} \langle \omega |T_{ab}| \omega \rangle k^a k^b \,
[c(v/ \lambda
)]^2 \, dv
\end{equation}
satisfies $\beta_c (k) > -\infty$. Generalized ANEC reduces (or, more
accurately, is strengthened) to ANEC when one
imposes the stronger condition:
$\beta_c (\gamma_\ast ) \geq 0$ for all weighting functions $c(x)$
as above. Note that generalized ANEC can be formulated equivalently
in the (perhaps more sensible) form of an inequality: namely, for all
$c(x)$ chosen as above,
\begin{equation}
\liminf_{\lambda \rightarrow \infty}
\int_{\gamma} \langle \omega |T_{ab}| \omega
\rangle k^a k^b \, [c(v/ \lambda
)]^2 \, dv \; \geq \; \beta_c (k) \; \; \; \; \; \; \; \forall \omega
\; ,
\end{equation}
where the expression $\beta_c (k)$ on
the right hand side is a state-independent lower bound for
the (weighted) ANEC integral on the left
hand side. In general (when generalized ANEC holds), this lower bound
(i.e., the precise expression of the 1-form $\beta_c$) will depend only
on the geometry of the spacetime $(\SM ,g)$ [as well, of course, as on
the null geodesic $\gamma$ and the weighting function $c(x)$]. If the
infimum over $\omega$ in Eq.\,(2) is achieved [for all $c(x)$]
by some Hadamard state $\omega_0$, so that
\[
\beta_c (k) =\liminf_{\lambda \rightarrow \infty}
\int_{\gamma} \langle \omega_0 |T_{ab}|
\omega_0 \rangle k^a k^b \, [c(v/ \lambda
)]^2 \, dv \;  ,
\]
then Eq.\,(3) can be written in the form of a
{\sl difference inequality} (see [3] and [1] on difference
inequalities):
\begin{equation}
\liminf_{\lambda \rightarrow \infty}
\int_{\gamma} (\, \langle \omega |T_{ab}| \omega
\rangle \, - \, \langle \omega_0 |T_{ab}| \omega_0 \rangle \, ) \,
k^a k^b \, [c(v/ \lambda
)]^2 \, dv \; \geq \; 0 \; \; \; \; \; \; \; \forall \omega \; .
\end{equation}
Conversely, if $\langle \omega | T_{ab}(x) | \omega \rangle$ satisfies a
difference inequality
\[
\liminf_{\lambda \rightarrow \infty}
\int_{\gamma} (\, \langle \omega |T_{ab}| \omega
\rangle \, - \, D_{ab} \, ) \,
k^a k^b \, [c(v/ \lambda
)]^2 \, dv \; \geq \; 0 \;
\; \; \; \; \; \; \forall \omega \]
such that the expression $D_{ab} k^a k^b $ is integrable
along $\gamma$, then
generalized ANEC [Eq.\,(3)] holds with $\beta_c (k)
\geq [c(0)]^2 \int_{\gamma} D_{ab} k^a k^b \, dv $.

Before I discuss the physical significance of generalized ANEC, let me
explain why this modified version of ANEC has a better chance of holding
generally in curved four-dimensional spacetime than the original
version. Recall the scaling argument given in the note added in proof to
[2] and discussed in more detail recently in [4]: Restrict attention,
for simplicity, to a massless Klein-Gordon scalar quantum
field $\phi$. Given an arbitrary
four-dimensional spacetime $(\SM ,g)$
in which $\phi$ satisfies ANEC along a null
geodesic $\gamma$, the scaling argument asks us to
consider the new spacetime $(\SM ,\kappa^2 g)$, where $\kappa >0$
is a {\sl constant} scale factor (in particular, in this
new spacetime the curve $\gamma \subset \SM$ is still a
null geodesic with the same affine parameter $v$).
To every Hadamard state $\omega$ of $\phi$ with two-point
function $\mu_\omega (x,x')$ on the original spacetime, there
corresponds a Hadamard state $\omega$ on the scaled spacetime
with two-point function $\kappa^{-2} \mu_\omega (x,x')$. [The massless
Klein-Gordon equation is invariant under scale transformations (whereas
the massive Klein-Gordon equation is scale-invariant
upto a rescaling of the mass); therefore the
function $\mu_\omega (x,x')$ remains a
bi-solution of the massless Klein-Gordon equation under
the scaling $g \rightarrow \kappa^2 g$. The overall scale factor
$\kappa^{-2}$ is introduced to keep $\mu_\omega$ in Hadamard form in the
new spacetime.] Normally, then, one would expect the
regularized expectation value $\langle \omega | T_{ab} | \omega
\rangle$ to simply scale as $\kappa^{-4}$ (because its definition involves
differentiating the two-point function twice
with respect to locally inertial
coordinates). However, according to the general renormalization
prescription for $\mu_\omega$, {\sl before} the differentiations and
the limit $x \rightarrow x'$ are carried out to evaluate
$\langle \omega |T_{ab}(x)| \omega \rangle$,
a locally constructed Hadamard distribution
$\mu_0 (x,x')$ needs to be subtracted from $\mu_\omega (x,x')$ to obtain
the regularized two-point function. It turns out that
this local Hadamard distribution
$\mu_0$ does not scale in the same simple way as $\mu_\omega$ under the
scaling $g \rightarrow \kappa^2 g$ of the metric, and this anomalous scaling
behavior of $\mu_0 (x,x')$ ends up contributing two
additional terms (apart from the simply
scaled term $\kappa^{-4} \langle \omega |T_{ab}|
\omega \rangle$) to the value of $\langle \omega |T_{ab}| \omega
\rangle$ in the scaled spacetime ([5]). These additional terms are
of the form \mbox{$a \, \kappa^{-4} \ln \kappa \,
\; ^{(1)} \! H_{ab} \; + \;
b \, \kappa^{-4} \ln \kappa \, \;^{(2)}\!H_{ab}$}, where
\mbox{$a$, $b$} are
dimensionless (in Planck units) constants
which have known universal values for each
fixed quantum field, and $^{(1)}\!H_{ab}$ and $^{(2)}\!H_{ab}$
denote the conserved local curvature terms
\begin{equation}
\, ^{(1)}\! H_{ab} \, \equiv \,  2 R_{;ab} \, + \, 2 R R_{ab} \, - \,
g_{ab}(2 \, \mbox{\raisebox{-.3ex}{$\Large\Box$}} R
+ \mbox{$1 \over 2$} R^2 ) \; ,
\end{equation}
and
\begin{equation}
\, ^{(2)}\! H_{ab} \, \equiv \, R_{;ab} \, - \,
\mbox{\raisebox{-.3ex}{$\Large\Box$}} R_{ab}
\, + \, 2 {R_a}^c R_{cb} - \mbox{$1 \over 2$} g_{ab} (\,
\mbox{\raisebox{-.3ex}{$\Large\Box$}} R
+ R^{cd} R_{cd} ) \; ,
\end{equation}
respectively. The precise numerical
values of the constants $a$ and $b$ depend only on the spin and
internal structure of the specific quantum field considered.
[Note that although this scaling behavior
of $\langle \omega |T_{ab}| \omega \rangle$
is closely related to the famous ambiguity in the
renormalization prescription, the constants $a$ and $b$ are
determined independently of this ambiguity. For most fields of
interest their values can be found in the literature
(see [5], p.\,1450 for a table of these constants for
various quantum fields; notice, however, that $a$ and $b$
in that table are given with respect to different
conserved curvature terms which are linear combinations of
$^{(1)}\! H_{ab}$ and $^{(2)}\! H_{ab}$). I will not need to specify the
exact values of $a$ and $b$ in this note; it will suffice to know
only the fact that in general these are constants
with absolute magnitudes of order $10^{-4}$ (in Planck units).]
$\,$It is now clear that if the curvature of the original spacetime is
sufficiently general so that the integrals $\int_\gamma \, ^{(1)} \! H_{ab}
k^a k^b \, dv$ and
\mbox{$\int_\gamma \, ^{(2)} \! H_{ab} k^a k^b \, dv$} are
non-vanishing, then by choosing the scale factor $\kappa$ appropriately
(note that the logarithm $\ln \kappa$ has indefinite sign)
it should be possible to find a spacetime $(\SM ,\kappa^2 g)$ in
which ANEC is violated along $\gamma$. Notice, however, the crucial
feature of the ANEC-violating term (proportional
to $\kappa^{-4} \ln \kappa$) disclosed
by this scaling argument: it is independent of the quantum \mbox{state
$\omega$}. Therefore, if as a functional of the quantum state
the ANEC integral along $\gamma$ is bounded from below in the original
spacetime (as would be the case if ANEC holds there),
with the greatest lower bound given by a 1-form $\beta$ as in
Eq.\,(1), then the only effect of the scaling
$g \rightarrow \kappa^2 g$ will be to shift this
lower bound $\beta$ down (or up) by an amount
proportional to $\kappa^{-4} \ln \kappa$
and the integrals of $^{(1)} \! H_{ab} k^a
k^b$ and $^{(2)} \! H_{ab} k^a k^b$ along $\gamma$. When the spacetime
$(\SM ,g)$ is asymptotically well-behaved
(so that its Ricci curvature falls off
appropriately at null infinity), these integrals are finite.
Consequently, {\sl if generalized ANEC holds along $\gamma$
in the asymptotically flat (more precisely, asymptotically empty)
spacetime $(\SM ,g)$, then
it holds in the scaled spacetime $(\SM ,\kappa^2 g)$ for any $\kappa >0$.}

In the remainder of this note I will argue that generalized ANEC,
although a much weaker constraint than the usual ANEC, has physical
significance quite similar to that of ANEC
in semiclassical gravity. I will make this argument by
discussing in turn the implications of generalized ANEC for
positivity of total energy, for singularity theorems, and for the
existence of static traversable wormhole solutions to the semiclassical
Einstein equations.

{\it Positivity of total energy.---} Recall the argument in Ref.\,[1]
leading to the Theorem in Sect.\,1 there. Instead of a $T_{ab}$
satisfying ANEC in the simple form Eq.\,(3) of [1], consider a quantum
stress-energy tensor $\langle \omega | T_{ab} | \omega \rangle$ which
satisfies generalized ANEC in the form
\begin{equation}
\int_{\gamma} \langle \omega |T_{ab}| \omega
\rangle k^a k^b \,
dv \; \geq \; \beta_\gamma (k) \; \; \; \; \; \; \; \forall \omega
\;
\end{equation}
along all complete null geodesics $\gamma$ in $(\SM ,g)$ [assume, in
other words, that $\langle \omega | T_{ab} | \omega \rangle$ is
integrable along each complete $\gamma$ and satisfies Eq.\,(3) above].
Consider a Cauchy surface $\Sigma$ and a subregion $S \subset \Sigma$ as
in [1], and modify the assumption (A1) in Sect.\,1 of\vspace{5pt} [1] to:
\newline (A1) For each fixed Hadamard state $\omega$, let
the subregion $S \subset \Sigma$ be chosen large enough such that
generalized ANEC [Eq.\,(7)] holds for
$\langle \omega | T_{ab} | \omega \rangle$ along
all null generators of the future horizon $H^{+}(S)$\vspace{5pt}.
\newline Assume also that the assumption (A2) holds as described in [1].
Then, using exactly the same arguments as in
the proof of the Theorem in Sect.\,1 of [1]
[\mbox{between} Eqs.\,(6) and (10) there], it follows that either the
total energy contained in $S$,
$\int_{S} \langle \omega | T_{ab} | \omega \rangle
n^{a}n^{b} \, d^3 \sigma$, is nonnegative, or, if this
total energy is negative,
then it is bounded from below by a lower bound which depends on
$\omega$ only weakly through the \mbox{choice of $S$}. More
precisely, it follows that
\begin{eqnarray}
& & \int_{S} \langle \omega | T_{ab} | \omega \rangle
n^{a}n^{b} \, d^3 \sigma \; \geq \; \nonumber \\
& & \min \left[ 0 \, , \; \sup_{\alpha} \left(
\left| \frac{1}{\Vert \nabla \alpha \Vert_{\,_S} } \right|
\int_{\gamma \in H^{+}(S)}
d^2 \Omega \; \beta_{\gamma} (- \nabla \alpha ) \right) \right]
\; \; \; \; \; \forall \omega \; ,
\end{eqnarray}
where the supremum on the right hand side is over all time functions
$\alpha$ which satisfy (for {\sl some} constants $\kappa$, $q >0$) the
conditions of assumption (A2), and the integral inside the $\sup_\alpha$
is over all null generators of the horizon $H^{+}(S)$,
evaluated with respect to the unique ``solid angle"-measure
$d^2 \Omega$ on the set of generators such that $d^2 \Omega \, dv
= d^3 \sigma$ [where $v$ is the affine parameter
along the generators, and $d^3 \sigma$ is the
canonical volume element of $H^{+}(S)$]. Consequently, just as ANEC
places a positivity constraint on the integrated energy density under
appropriate assumptions, so also generalized ANEC places, under
similar assumptions, an essentially state-independent
(in general negative) lower bound on the same quantity. (Note
that in general the quantum state $\omega$ determines
exactly how ``large" the region $S \subset \Sigma$ needs to be chosen,
and this is the only reason the lower bound might depend on $\omega$.)

{\it Singularity theorems.---} To illustrate the relevance
of generalized ANEC for singularity theorems and other global results of
classical General Relativity, recall the Proposition proved in Sect.\,2
of [2], which uses the constraint on the Ricci tensor
imposed by ANEC and the Einstein equations to demonstrate a focusing
lemma for null geodesics; a result of the kind which
constitute the key ingredient in the proof of global
results such as singularity theorems.
\mbox{A straightforward} reworking of the
argument in the proof described in Ref.\,2 [given between Eqs.\,(4) and
(9) there] directly demonstrates the following variation
\vspace{5pt}of that Proposition:
\newline {\it Proposition}: Let $p$ be a point on a complete null
geodesic $\gamma (v)$. Assume that $\gamma$
satisfies the following property:
For the specific choice of the weighting function $c(x)$ [see the
formulation Eq.\,(3) of generalized ANEC above] given by
\begin{eqnarray}
c(x) = c_1 (x) & = &
(1 -  x^{2} )^{2}  \, , \; \; \; \; |x| < 1 \; ,
\nonumber \\
c_1 (x) & = & 0 \, , \; \; \; \; \; \; \; \; \; \;
\; \; \; \; \; \; \, |x| \geq 1 \; ,
\end{eqnarray}
the weighted
average of the Ricci tensor $R_{ab}$ along $\gamma$ obeys the
inequality
\begin{equation}
\liminf_{ \lambda \rightarrow \infty } \int_{0 }^{ \infty }
R_{a b} k^{a} k^{b} \,
[c_1 ( v/ \lambda )]^{2} \, dv \, \geq \, \beta_1 (k) \; ,
\end{equation}
where $\beta_1 (k) > -\infty$ (and where
$ v=0 $ at $p$). [If generalized ANEC (together
with semiclassical Einstein's equations) holds,
then with $\beta_1 \equiv 4 \pi \beta_{c_1}$
this condition must hold in every Hadamard state of the quantum
field for at least
one direction along $\gamma$ from $p$.]
Consider a null geodesic congruence
containing $\gamma$ whose expansion $\theta (v)$ along $\gamma$
satisfies, initially at the point $p$, the inequality
\begin{equation}
\theta (0) \, \leq \, \beta_1 (k) \;
\end{equation}
(note that under a rescaling of the affine parameter
the expansion $\theta$ scales in the same way as the tangent
vector $k$, so this
inequality is independent of the choice of affine parameter).
Then, either $\theta$ vanishes
identically along $\gamma$,
or there exists a finite $v_{0} > 0$ at
which $\lim_{ v \rightarrow v_{0} }
\theta (v) = - \infty $\vspace{5pt}.
\newline Therefore, independently of which quantum state the
field is in, generalized ANEC guarantees the refocusing of
a null geodesic congruence if the initial convergence is sufficiently
nonpositive, or, in other words (assuming \mbox{$\beta_1 \leq 0$}), if the
initial convergence is more negative than the
amount of ANEC violation allowed by generalized ANEC. As
was also the case with ANEC
(see the last paragraph on p.\,405 of Ref.\,[2]), a proof can also probably
be given
that if generalized ANEC holds along a complete null geodesic $\gamma$,
and if $\gamma$ satisfies the null generic
condition such that the maximum magnitude
of the quantity
$k^a k^b k_{[c} R_{d]ab[e} k_{f]}$
(which enters the formulation of the generic condition) is sufficiently
large compared to the magnitude of ANEC violation allowed by
generalized ANEC [cf.\ Eq.\,(11)], then $\gamma$ must contain a pair of
conjugate points.

{\it Traversable wormholes.---} A widely applicable
generalized ANEC theorem would place a significant a priori constraint
on possible solutions to the semiclassical Einstein equations.
Namely, assume that such a theorem---to the effect that generalized ANEC
holds along (certain) complete null geodesics $\gamma$
in every asymptotically empty spacetime, with a geometric,
state-independent lower bound $\beta_\gamma$---were available. Then,
given any spacetime $(\SM ,g)$, one could compute for each
specified $\gamma \subset \SM$ the
quantity $\beta_\gamma$ in the geometry of $(\SM ,g)$, and compare
the result with the
ANEC integral of the Einstein tensor ${1\over{8 \pi}} G_{ab}$ along the
same null geodesic. If the comparison fails to satisfy the generalized
ANEC inequality [Eq.\,(3)] for at least some $\gamma$, then there cannot
exist any Hadamard quantum state which would make $(\SM ,g)$ a
self-consistent semiclassical solution of the Einstein equations;
in other words, the spacetime $(\SM ,g)$ would be
ruled out by quantum field theory (at least
with the specific quantum fields for which a detailed analysis of the ANEC
integrals can be carried out). A nice illustration of these
ideas is provided by static (spherically symmetric) traversable
wormhole spacetimes. Such a wormhole has
topology $S^2 \times \BR^2$, and a metric of the general form
\begin{equation}
g \, = \, -e^{2 \Phi (l)} \, dt^2 \, + \, dl^2 \, + \,
r(l)^2 (d \theta^2 + \sin^2 \theta \, d \phi^2 ) \; ,
\end{equation}
where the radial coordinate $l$ ranges from $- \infty$ (on one
asymptotic region) to $+ \infty$ (on the other). For there
to be no event horizons (hence for the wormhole to be traversable),
$\Phi$ needs to be finite everywhere. For asymptotic
flatness, it is necessary that
as $l \rightarrow \pm \infty$ (more precisely,
for $|l| \gg r_0$)
\begin{equation}
r(l) \, \simeq \, |l| \, - \, M \, \ln \left( \frac{|l|}{r_0} \right) \; ,
\; \; {\rm and} \; \; \Phi (l) \, \simeq \, - \frac{M}{|l|} \; ,
\end{equation}
where $r_0$ is the radius of the
wormhole's ``throat" (where $l$ vanishes), and $M$ is the wormhole's
mass. Throughout my discussion here I will assume that $r_0
\sim 2 M$ (which should be the case if
as seen from infinity the wormhole is
indistinguishable from an astrophysical object);
as a result, the class of wormholes I will consider
is parametrized (essentially) by one variable: the wormhole mass $M$.
For more details on wormholes see the
discussion in [6]; for a more up-to-date account
(including a discussion of the more recent work on ANEC) see [7].

The spacetime given by Eqs.\,(12--13) [with everywhere regular $\Phi (l)$]
violates ANEC along all its radial null geodesics:
a straightforward computation of the Einstein tensor followed by an
integration by parts reveals that
\begin{equation}
V \, \equiv \, \frac{1}{8 \pi} \int_{\gamma} G_{ab} k^a k^b \, dv
\, = \, -\frac{1}{4 \pi} \int_{-\infty}^{\infty} e^{-\Phi}
\left( \frac{r'}{r} \right)^2 \, dl \;
\end{equation}
along any radial null geodesic $\gamma$ (note that the affine parameter
$v$ along $\gamma$ can be chosen to be any positive constant times $\int
e^{\Phi} \, dl \, $; I will choose this constant to be unity throughout so
that $dv = e^{\Phi} \, dl $ and $k=e^{-2\Phi}\, \partial /\partial t
\, + \, e^{-\Phi} \, \partial /\partial l $). Here and in what follows a
prime ($'$) denotes differentiation with respect to the radial
coordinate $l$. Substituting Eq.\,(13) in the last integral of Eq.\,(14)
(and carrying out the integration only over $|l| \geq 2 M$) gives
\begin{equation}
V \, \approx \, -\frac{67}{384 \pi} \frac{1}{M} \; .
\end{equation}
Can this ANEC violation $V$ necessary to maintain a traversable wormhole
be supported by a quantum stress-energy tensor?
Consider a massless Klein-Gordon field on the wormhole
spacetime (the answer is not likely to
depend significantly on the spin or internal structure of the
field). Assume that generalized ANEC holds
along the radial null geodesics of the wormhole. Whether or not this
assumption is true remains to be shown; however, the scaling
argument sketched above combined with the
known ANEC theorems ([2]) in two and four dimensions suggest that it
is likely to be true (note that the wormhole's
radial null geodesics are complete
and achronal). Proceeding with the assumption that generalized ANEC
holds, how can we guess the form that
the (finite) ANEC lower bound $\beta (k)$
is likely to take
along the radial null geodesics of the
wormhole spacetime? One way to approach this question
is to look closely at the scaling behavior of the wormhole metric
Eq.\,(12). In general, a scaling $g \rightarrow \kappa^2 g$
of Eq.\,(12) leads to a
new wormhole metric $\bar{g}=\kappa^2 g$
for which the metric functions $\Phi$ and
$r$ are given by
\begin{equation}
\bar{\Phi}(l) \, = \, \Phi ( l/\kappa ) \; ,
\; \; {\rm and} \; \; \bar{r} (l) \, = \, \kappa \, r (
l/\kappa ) \; ,
\end{equation}
where a bar over a symbol indicates that the corresponding
quantity refers to the
scaled spacetime with metric $\bar{g}=\kappa^2 g$. From Eq.\,(14) it
follows quite generally that $\bar{V}=\kappa^{-1} V$, and from
Eq.\,(13) it follows that $\bar{r}_0 = \kappa r_0$ and [consistent with
Eq.\,(15)] $\bar{M}=\kappa M$. From the scaling argument I described
above for a general spacetime, it follows that
\begin{equation}
\overline{\beta (k)} \, = \, \frac{1}{\kappa^3} \, \beta (k) \, + \,
\frac{\ln \kappa}{\kappa^3} \,
\int_{\gamma} \left( a \; ^{(1)}\!H_{ab} k^a k^b
\, + \, b  \; ^{(2)}\!H_{ab} k^a k^b
\right) \, dv
\; ,
\end{equation}
where $^{(1)}\!H_{ab}$ and $^{(2)}\!H_{ab}$ are
the conserved curvature terms given by Eqs.\,(5) and
(6), respectively. Computer algebra systems ([8]) make the computation
of these higher curvature terms easier:
for the wormhole metric [Eq.\,(12)] I find
\begin{eqnarray}
^{(1)}\!H_{ab} k^a k^b \, dv
& = &
-  8 \left[ {\rm TD} - \left( e^{-\Phi} \right)''
\Phi '' - 3 \left( r' e^{-\Phi} \right)' \frac{r' \Phi'}{r^2}
\right. \nonumber \\
& + & e^{-\Phi} \left( \Phi'''' + \frac{r'}{r} \Phi' \Phi''
 -
\frac{(r')^2}{r^2} \Phi'' + \frac{r'}{r} (\Phi')^3
\right. \nonumber \\
& + & 3 \frac{(r')^2}{r^4} [ 1 + (r')^2 ]
+  \frac{r''''}{r}
-\frac{(r'')^2}{r^2}
- 6 \frac{r''}{r^3} (r')^2
\nonumber \\
& - & \left. \left. \frac{r''}{r^2}
r' \Phi' + 4 \frac{(r')^3}{r^3} \Phi' - 2 \frac{r''}{r} (\Phi')^2
\right) \right] \, dl \;,
\end{eqnarray}
and
\begin{eqnarray}
^{(2)}\!H_{ab} k^a k^b \, dv
& = &
-  2 \left[ {\rm TD} - 2 \left( e^{-\Phi} \right)''
\Phi '' - 7 \left( r' e^{-\Phi} \right)' \frac{r' \Phi'}{r^2}
\right. \nonumber \\
& + & e^{-\Phi} \left( \Phi'''' + 5 \frac{r'}{r} \Phi' \Phi''
+ \frac{r'}{r} \Phi''' + \frac{r'''}{r} \Phi'
\right. \nonumber \\
& + & 2 \frac{(r')^2}{r^4} [ 1 + 2 (r')^2 ]
- 2 \frac{(r'')^2}{r^2}
- 6 \frac{r''}{r^3} (r')^2
\nonumber \\
& + & \left. \left.
4 \frac{(r')^3}{r^3} \Phi' - \frac{r''}{r} (\Phi')^2
\right) \right] \, dl \;,
\end{eqnarray}
where the symbols TD denote total derivative terms of the form $dF[\Phi,
r]/dl$ with $\lim_{l \rightarrow \pm \infty} F = 0$ under the boundary
conditions Eqs.\,(13). A calculation similar to the derivation of
Eq.\,(15) from Eq.\,(14) gives, when applied to Eqs.\,(18)--(19),
\begin{equation}
\int_{\gamma} \, ^{(1)}\!H_{ab} k^a k^b \, dv \, \approx \,
\frac{3}{128} \frac{-1221 + 2048 \ln (2) }{M^3} \, \approx \,
4.65 \frac{1}{M^3} \; ,
\end{equation}
and
\begin{equation}
\int_{\gamma} \, ^{(2)}\!H_{ab} k^a k^b \, dv \, \approx \,
\frac{1}{2520} \frac{-28117 + 47040 \ln (2)}{M^3} \, \approx \,
1.78 \frac{1}{M^3} \; .
\end{equation}
Now let $B(M)$ denote the value of the quantity $\beta (k)$ for a
wormhole spacetime [Eqs.\,(12)--(13)] of mass $M$ (recall
$r_0 \sim 2 M$). Note that, for simplicity, I have been ignoring
in this discussion the
more accurate version Eqs.\,(2)--(3) of generalized ANEC;
more precisely, I have assumed (and will continue
to assume) that for a general weighting
function $c(x)$ it holds that $\beta_c (k) = [c(0)]^2 \beta (k)$.
Combining Eqs.\,(20)--(21) with Eq.\,(17) and recalling that
$a \sim b \sim 10^{-4}$, I deduce the relation
\begin{equation}
B(\kappa M) \, = \, \frac{1}{\kappa^3} B(M) \, + \,
\frac{\ln \kappa}{\kappa^3} \, \frac{10^{-4} c}{M^3} \; ,
\end{equation}
where $c$ is a numerical constant with $|c| \sim 1$. It is reasonable to
guess (and this is the only ``guessing" involved in the present
argument) that $|B(1)| \sim 1$; i.e., that the value of $B(M)$ for a
Planck-mass ($M \sim 1$) wormhole is (in absolute magnitude) of order
unity, that is, of Planck size (in Planck units). Then Eq.\,(22) gives
\begin{equation}
B(M) \, \approx \, \frac{1}{M^3} \left( c_1 + 10^{-4} c_2 \ln M \right)
\; ,
\end{equation}
where $c_1$ and $c_2$ are constants with $|c_1 | \sim |c_2 | \sim 1$.
Clearly, for reasonable $M$ the second term in parenthesis in Eq.\,(23)
is negligible compared to the first; hence $|B(M)| \sim 1/ M^3$.
In order to have the ANEC violation $V$ [Eq.\,(15)] supportable by the
renormalized stress-energy tensor, it is necessary that $B(M)
\leq V$, which is only possible if $|B(M)| \geq |V|$, which implies
$M^2 \leq c_3$, where $c_3 \sim 1$. Therefore, {\sl if generalized ANEC
holds, a quantum Klein-Gordon field can support the ANEC violation
necessary for a traversable wormhole only if the wormhole has Planck
mass or less; in other words, all traversable wormholes
except possibly those of Planck size are ruled
out by quantum field theory.}

It is important to keep in mind that this conclusion rests entirely on
the assumption that generalized ANEC holds along the
radial null geodesics of the wormhole spacetime. If it
can be shown that the ANEC integral
along these geodesics is unbounded from below as a functional of the
quantum state, then no amount of ANEC violation can be ruled out by
quantum field theory; in particular, there might exist states
in which the violation Eq.\,(15) necessary for a
macroscopic traversable wormhole is supported
by the expected quantum stress-energy tensor.
Even if generalized ANEC could be shown to hold by
indirect methods, a rigorous computation of the lower bound $\beta (k)$
will be necessary to demonstrate that
it indeed has the behavior described in
Eqs.\,(22)--(23). [It is plausible to conjecture that if generalized ANEC
holds, then it can be put in the form Eq.\,(4),
with the ``minimum-ANEC-integral" state $\omega_0$ being given
by the standard, isometry-invariant
vacuum for the wormhole metric.]$\,$ Also, my conclusion is based on
the analysis of a
simple one-parameter family of wormholes; the
general criterion to decide which wormhole metrics Eq.\,(12) are
allowed in semiclassical gravity is bound to be much more complicated.
For example, one could choose the function $r(l)$ to be very slowly
varying so as to make the integral in Eq.\,(14) as small (in absolute
value) as the microscopic ANEC violation allowed by quantum
field theory. It appears that a macroscopic wormhole of this
kind (with extremely small radial curvature)
could be allowed even when generalized ANEC holds; however, such a
spacetime would look more like a constant-radius $S^2 \times \BR^2$
universe than a wormhole joining two asymptotically flat regions.
Finally, although the main conclusion is almost certainly independent
of the spin and internal structure of the specific Klein-Gordon field,
the argument above deals only with the question of maintaining
a wormhole using free (non-interacting) quantum fields. A plausible
configuration for wormhole maintenance based on the Casimir effect
(where an electromagnetic field providing the
negative Casimir energy interacts with the
matter fields in the conducting plates which trap it in the Casimir
vacuum state) was outlined in Ref.\,[6]. To rule out
macroscopic configurations of that sort, one would need
a generalized ANEC theorem applicable
to interacting fields. Nevertheless, it is difficult to see how
interactions could induce a violation of generalized ANEC
if such a theorem holds generally in
non-interacting quantum field theory.

Similar conclusions about the constraint imposed by generalized ANEC
on solutions of semiclassical Einstein equations can be reached
for spacetimes more general than wormholes. For example, in Ref.\,[9] it
is shown that when the classical null energy condition holds, a region
with non-trivial topological structure cannot be visible from infinity
in an asymptotically flat spacetime. Combining the
proof of this result with arguments similar to above, it can be shown
that if generalized ANEC holds along complete null geodesics $\gamma$ in
an asymptotically flat spacetime, with a microscopic (i.e., proportional
to $\hbar$) ANEC lower bound $\beta_\gamma$, then a nontrivial
topological structure can be visible from infinity only if its
spacetime curvature is of the Planck size.

{}~~~~~~~

{\bf \noindent Acknowledgements}

The possibility that the scaling argument of our earlier paper ([2])
may not allow macroscopic violations of ANEC while permitting
microscopic violations was suggested to me by R.\ Wald in Santa Barbara
two years ago. This research was carried out at the Jet
Propulsion Laboratory, California Institute of Technology, and was
sponsored by the NASA Relativity Office and by the National Research
Council through an agreement with the National Aeronautics and Space
Administration.

\newpage

\begin{center}
{\bf REFERENCES}
\end{center}

\noindent{\bf 1.} U.\ Yurtsever, JPL-Caltech Preprint (November, 1994).

\noindent{\bf 2.} R.\ Wald and U.\ Yurtsever, Phys.\ Rev.\ D {\bf 44},
403 (1991).

\noindent{\bf 3.} L.\ H.\ Ford and T.\ A.\ Roman, Tufts University
Preprint (1994).

\noindent{\bf 4.} M.\ Visser, Washington University Preprint (1994).

\noindent{\bf 5.} G.\ T.\ Horowitz, Phys.\ Rev.\ D {\bf 21},
1445 (1980).

\noindent{\bf 6.} M.\ S.\ Morris, K.\ S.\ Thorne and U.\ Yurtsever,
Phys.\ Rev.\ Letters {\bf 61}, 1446 (1988).

\noindent{\bf 7.} K.\ S.\ Thorne, in {\it
Proceedings of the 13th
International Conference on General Relativity and Gravitation},
edited by R.\ J.\ Gleiser, C.\ N.\ Kozameh, and O.\ M.\ Moreschi,
Institute of Physics Publishing, Bristol, England, 1993.

\noindent{\bf 8.} P.\ Musgrave, D.\ Pollney and K.\ Lake, ``GRTensor II
for Maple V Release 3," Queen's University, Kingston, Ontario, Canada
(1994).

\noindent{\bf 9.} J.\ Friedman, Phys.\ Rev.\ Letters {\bf 71},
1486 (1993).

\end{document}